\documentclass[10pt,letterpaper]{article}
\usepackage{opex3}
\usepackage{cite}
\usepackage{graphicx}
\usepackage{epstopdf}

\begin{document}

\title{Spontaneous Rayleigh seeding of stimulated Rayleigh scattering in high power fiber amplifiers}

\author{Arlee V. Smith$^*$ and Jesse J. Smith}

\address{AS-Photonics, 6916 Montgomery Blvd., Ste. B8, Albuquerque, NM 87109, USA}

\email{$^*$arlee.smith@as-photonics.com}

\begin{abstract}
{We estimate the Stokes wave starting power for stimulated thermal Rayleigh scattering (STRS)  produced by thermal fluctuations in the fiber core that transiently alter the refractive index profile in the core. A transverse temperature gradient creates a transverse refractive index gradient via the thermo optic effect, and if the fluctuation frequency lies in the STRS gain band, it can couple light from mode LP$_{01}$ to LP$_{11}$ to seed STRS. This spontaneous Rayleigh seed may be stronger than the quantum background and may affect the mode instability thresholds of fiber amplifiers. This new seed estimate can be incorporated in STRS models to improve threshold calculations.}
\end{abstract}


\section{Introduction}

Several authors have demonstrated that stimulated thermal Rayleigh scattering (STRS) in high power fiber amplifiers leads to high gain for light in higher order modes and can account for observations of modal instability in such amplifiers\cite{smith1,smith2,hansen1,hansen2,dong,ward}. Just as for other stimulated scattering processes such as Raman (SRS) or Brillouin (SBS), precise calculations of threshold powers require a good estimate of the starting light level in the Stokes wave. In SRS quantum noise of the Stokes optical field usually serves as the starting light, while in SBS thermally induced acoustic waves in the fiber core usually initiate the process \cite{boyd,agrawal}. Here we calculate the strength of inelastic spontaneous thermal Rayleigh scattering and show that it can initiate STRS even if the signal and pump are unmodulated. The spontaneous scattering is caused by thermal fluctuations in the temperature profile in the core which creates refractive index fluctuations via the thermo optic effect. The strength of the spontaneously scattered light is often greater than the quantum background that has been invoked in most previous estimates of the maximum STRS threshold power\cite{smith1,smith2,hansen1,hansen2,dong}.

In most current fiber amplifier systems it is likely that amplitude modulation of the pump or seed light supplies an STRS Stokes seed that is much stronger than both the quantum noise and the spontaneous thermal Rayleigh scattering that is the subject of this study\cite{smith4,hansen2,smith5}. However, future progress in increasing STRS thresholds will require an understanding of the fundamental limits imposed by these noise sources as well as the limits imposed by technical amplitude noise on the pump and seed light. Seed amplitude noise can be minimized by using a nearly quantum limited seed such as a stabilized nonplanar ring Nd:YAG laser\cite{harb} or other well stabilized source. Progress in amplitude and wavelength stabilizing the pump diodes will also be necessary if the fundamental limits are to be approached. It is important to note that only amplitude modulation of the injected light affects the STRS threshold. Phase modulation such as that used to suppress SBS does not contribute\cite{hansen2}. 

We will first present a brief, inexact description of the spontaneous thermal Rayleigh scattering process, followed by a more exact theory. Thermodynamic fluctuations produce transient variations in the energy contained in a given volume in the vicinity of the fiber core. This energy variation is largely in the form of temperature fluctuations. These fluctuations have a frequency and spatial spectrum which, if it matches the profile required to shift light from the fundamental mode into the higher order mode with the proper frequency shift, can serve to seed the STRS process. The second moment of the temperature fluctuations, averaged over a volume $V$ in contact with a thermal reservoir at temperature $T$, has the time averaged value\cite{callen} of $(k_BT^2)/(\rho C V)$, with the quantities for silica listed in Table 1. Calculating the portion of this fluctuation that is responsible for seeding the STRS process is the main topic of this report. To perform the calculation we modify the procedure developed by Gorodetsky and Grudinin\cite{gorodetsky} to compute thermally induced phase noise in microsphere resonators. The calculations in this paper supplant our earlier estimate\cite{smith4} of the upper limit of seeding by this process. That estimate was based on a null search of the literature for temperature dependent spontaneous Rayleigh scattering.
\begin{table}[htb]
\renewcommand{\arraystretch}{1.3}
\centering\caption{Silica parameters}
\begin{tabular}{c|c|c}\hline 
Boltzmann constant     &      $k_B$      &     $1.38\times 10^{-23}$ J$\;$/$\;$K\\
density                &      $\rho$     &      2201 kg$\;$/$\;$m$^3$\\
thermo optic coeff.    &      $\alpha=dn/dT$ &  $1.2\times 10^{-5}$ K$^{-1}$\\
specific heat          &      $C$        &      702 J$\;$/$\;$kg$\cdot$K\\
thermal conductivity   &      $\kappa$   &      1.38 W$\;$/$\;$m$\cdot$K\\ 
thermal diffusivity    &      $D=\kappa/\rho C$ &  $8.9\times 10^{-7}$ m$^2$$\;$/$\;$s\\
\hline
\end{tabular}
\end{table}

\section{Initial estimate of spontaneous scattering rate}

The thermal energy contained in a volume $V$ of silica is
\begin{equation}
U=\rho C V T
\end{equation}
and the time averaged second moment of the energy fluctuation in $V$ in contact with a thermal reservoir is\cite{callen}
\begin{equation}
<(\delta U)^2>=k_BT^2\frac{\partial U}{\partial T}=k_BT^2\rho C V.
\end{equation}
Assuming the energy fluctuations are predominantly in the form of temperature fluctuations, the second moment of the temperature fluctuation, averaged over volume $V$, is 
\begin{equation}\label{eq.dt1}
<(\overline{\delta T})^2>=\frac{k_BT^2\rho C V}{\rho^2 C^2 V^2}=\frac{k_BT^2}{\rho C V},
\end{equation}
where $\overline{\delta T}$ is the temperature deviation from $T$, averaged over the volume of interest. Using the tabulated values for $k_B$, $\rho$, and $C$ gives
\begin{equation}\label{eq.dt2}
<(\overline{\delta T})^2>=(8.92\times 10^{-30}\mbox{ m$^3$})\;\;\frac{T^2}{V}.
\end{equation}

Coupling light from the fundamental mode LP$_{01}$ to the two-lobed mode LP$_{11}$ requires a refractive index profile with a tilt across the fiber core. We will assume this tilt is due to a temperature slope across the core. If we divide the core into two semicircular cylinders and consider them to be the two volumes exchanging energy and causing the fluctuating temperature slope, the frequency spectrum of the fluctuations should peak near the inverse thermal diffusion time over the core radius. This frequency is similar to the frequency shift for maximum STRS gain\cite{smith3}. 

When one half of the core loans thermal energy to the other half in a thermal fluctuation, the phase shift between light traveling in the two core halves over the length of one beat between the modes is roughly
\begin{equation}
\Delta \phi=k_{\circ}\; L_B\; \alpha\; \overline{\delta T}
\end{equation}
where $L_B$ is the beat length, $k_{\circ}$ is the vacuum propagation constant, and for $\overline{\delta T}$ we use the square root of $<(\overline{\delta T})^2>$. Then using Eq. (\ref{eq.dt2}) gives
\begin{equation}\label{eq.dphi}
\Delta \phi= 3\times 10^{-15}\;k_{\circ}\; \alpha\;T\;\frac{L_B}{V^{1/2}}
\end{equation}

Because the ratio $L_B/V^{1/2}$ is nearly independent of the fiber core diameter, the phase shift $\Delta \phi$ per beat length is as well. For example, for a fiber with a numerical aperture of 0.054 this ratio has a value close to 3000 m$^{-1/2}$ for core diameters ranging from 20 to 80 $\mu$m. This nearly constant phase shift per beat length provides a handy measure of scattering strength. Using approximate values in Eq. (\ref{eq.dphi}) with $T=300$ K and $\lambda=1060$ nm gives
\begin{equation}
\Delta \phi\approx 2\times 10^{-7}\mbox{ radian}.
\end{equation}
This phase shift of 0.2 $\mu$rad after one beat length imposed on the LP$_{01}$ light in the two core halves can be decomposed into a strong LP$_{01}$ field combined with a weak LP$_{11}$ field, yielding a power ratio of $4\times 10^{-14}$ between the light in the LP$_{11}$ and fundamental modes. For 10 W of injected power in LP$_{01}$ ($P_{01}$) the Stokes starting power would thus be around 0.4 pW or $10^3-10^4$ times larger than the effective quantum noise power. 

However, this estimate is based on interpreting the thermal fluctuations as concentrated at a single frequency. In reality the speed of the fluctuations is limited by thermal diffusion and the size of the fiber core. These are the same factors that determine the frequency of maximum gain, $\omega_M$ for the STRS process. An educated guess is that the fluctuations scatter light over a frequency range $(-2\omega_M<\omega<+2\omega_M$). If we also assume that only scattered light within $\pm5\%$ of $\omega_M$ can affect the threshold power of STRS, the portion of the scattered power that can act as a seed is reduced by a factor of 40 to ($1\times 10^{-15}\times P_{01}$).

This treatment provides only an order of magnitude estimate of the seed power. A more precise calculation based on similar concepts is presented in the following sections.

\section{Improved estimate of spontaneous scattering rate}

\subsection{Coupled mode equations}

From coupled mode treatments of STRS\cite{ward} the coupling between modes LP$_{01}$ and LP$_{11}$ due to a temperature profile $T(x,y,z,t)$ is
\begin{equation}\label{eq.fieldcouple}
i\frac{\partial a_{11}}{\partial z}=k_{\circ}\alpha\biggl(\int \hat{E}_{01}T(x,y,z,t)\hat{E}_{11}dxdy\biggr)\;a_{01}\;e^{-i\Delta \beta z}
\end{equation}
where $a_{11}$ is the amplitude of the LP$_{11}$ field and $a_{01}$ is the amplitude of the LP$_{01}$ field, $\hat{E}_{01}$ is the LP$_{01}$ field normalized so $\int |\hat{E}_{01}|^2 dxdy=1$, and $\hat{E}_{11}$ is the same for LP$_{11}$. We will compute a fluctuational value of quantity $Q$ which is proportional to the coupling coefficient of Eq. (\ref{eq.fieldcouple}), 
\begin{equation}
Q(t)=\int\hat{E}_{01}(x,y)\hat{E}_{11}(x,y)\sin (2\pi z/L_B)\delta T(x,y,z,t)d{\bf r},
\end{equation}
where $\delta T$ is the antisymmetric part of the temperature deviation from the spatially averaged core temperature $T$. The sine factor is included to account for the $z$ dependence of the modal interference over one half the modal beat length $0<z<L_B/2$. This is also the range  for the $z$ integration in Eq. (9). We chose this range for convenience. Other ranges could be used but will yield almost the same value for the rate of power transfer per length of fiber.

\subsection{Computing $<|Q(t)|^2>$}

From here on we will use $T$ to represent $\delta T$ for simplicity. The average core temperature will be $T_{\circ}$. We assume the temperature fluctuations obey the thermal diffusion equation
\begin{equation}\label{eq.rt}
\frac{\partial T({\bf r},t)}{\partial t}=D\nabla^2 T({\bf r},t)+F({\bf r},t)
\end{equation}
where $F({\bf r},t)$ term is a fluctuating source, and $D$ is the thermal diffusivity of silica. Equation (\ref{eq.rt}) can be Fourier transformed to $\{{\bf k},\omega\}$ space and solved for $\widetilde{T}({\bf k},\omega)$,
\begin{equation}\label{eq.tkw}
\widetilde{T}({\bf k},\omega)=\frac{\widetilde{F}({\bf k},\omega)}{Dk^2+i\omega}.
\end{equation}
Fourier transforming back to $\{{\bf r},t\}$ space yields
\begin{equation}\label{eq.trt}
T({\bf r},t)=\frac{1}{(2\pi)^4}\int\frac{\widetilde{F}({\bf k},\omega)}{Dk^2+i\omega}\;e^{\;i(\omega t+{\bf k\cdot r})}\;\;d{\bf k}\;d\omega.
\end{equation}
It can be shown\cite{gorodetsky} that both Eq. (\ref{eq.dt1}) and the fluctuation-dissipation theorem are satisfied if $<\widetilde{F}({\bf k},\omega)\widetilde{F}^*({\bf k}^{\prime},\omega^{\prime})>$ has the form 
\begin{equation}\label{eq.ffstar}
<\widetilde{F}({\bf k},\omega)\widetilde{F}^*({\bf k}^{\prime},\omega^{\prime})>=\frac{2Dk_B}{\rho C}\;k^2\;T_{\circ}^2\;(2\pi)^4\;\delta({\bf k-k^{\prime}})\delta(\omega-\omega^{\prime}).
\end{equation}
To save space we rewrite $Q(t)$ in the form
\begin{equation}\label{eq.Q}
{Q}(t)=\int T({\bf r},t){\cal{E}}({\bf r})\; d{\bf r}
\end{equation}
where ${\cal{E}}$ is
\begin{equation}
{\cal{E}}({\bf r})=\hat{E}_{01}(x,y)\hat{E}_{11}(x,y)\sin(2\pi z/L_B).
\end{equation}
Using Eq. (\ref{eq.trt}) in Eq. (\ref{eq.Q}) gives
\begin{equation}
Q(t)=\frac{1}{(2\pi)^4}\int\frac{\widetilde{F}({\bf k},\omega)}{Dk^2+i\omega}\;{\cal{E}}({\bf r})\;e^{\;i(\omega t+{\bf k\cdot r})}\;\;d{\bf k}\;d\omega \;d{\bf r}.
\end{equation}
The Wiener-Khinchin theorem states that the spectral power of $<|Q(t)|^2>$ is equal to the Fourier transform of the auto correlation of $Q(t)$,
\begin{equation}
S(\omega^{\prime\prime})=\int\biggl[\int <Q(t)Q^*(t+t^{\prime\prime})> \;dt\biggr]\;e^{-i\omega^{\prime\prime}t^{\prime\prime}} \;dt^{\prime\prime}.
\end{equation}
\begin{equation}
S(\omega^{\prime\prime})=\frac{1}{(2\pi)^8}\int\frac{<\widetilde{F}({\bf k},\omega)\;\widetilde{F}^*({\bf k}^{\prime},\omega^{\prime})>}{(Dk^2+i\omega)(Dk^{\prime 2}-i\omega^{\prime})}\;\;{\cal{E}}({\bf r})\;\;{\cal{E}}^*({\bf r^{\prime}})\;e^{\;i(\omega t-\omega^{\prime}t-\omega^{\prime}t^{\prime\prime}-\omega^{\prime \prime}t^{\prime\prime}+{\bf k\cdot r-k^{\prime}\cdot r^{\prime}})}\;d\Omega 
\end{equation}
where $\Omega$ represents integrations over the variables $(tt^{\prime\prime}{\bf k}{\bf k^{\prime}}\omega\omega^{\prime}{\bf r r^{\prime}})$. Using the expression for\linebreak$<\widetilde{F}\widetilde{F}^*>$ from Eq. (\ref{eq.ffstar}) gives
\begin{equation}\label{eq.sw4}
S(\omega)=\frac{1}{(2\pi)^4}\frac{4Dk_BT_{\circ}^2}{\rho C}\int \frac{k^2}{D^2k^4+\omega^2}\;{\cal{E}}({\bf r})\;\;{\cal{E}}^*({\bf r^{\prime}})e^{i(\bf k\cdot r-k\cdot r^{\prime})}d{\bf k}\;d{\bf r}\;d{\bf r^{\prime}}.
\end{equation}
We define $G({\bf k})$ as the spatial Fourier transform of ${\cal{E}}({\bf r})$,
\begin{equation}
G({\bf k})=\int{\cal{E}}({\bf r})\;e^{i{\bf k}\cdot {\bf r}}\;d{\bf r}
\end{equation}
to write Eq. (\ref{eq.sw4}) as
\begin{equation}\label{eq.sw}
S(\omega)=\frac{1}{(2\pi)^4}\frac{4k_BT_{\circ}^2}{\rho C}\int \frac{Dk^2}{D^2k^4+\omega^2}\;|G({\bf k})|^2d{\bf k}.
\end{equation}
Because $Q$ is the temperature profile that couples the two modes, it has a limited spatial extent that we will call $V_{\rm eff}$. To find $V_{\rm eff}$ we integrate $S(\omega)$ over frequency to find the total power of $<|Q|^2>$ to be
\begin{equation}
<|Q|^2>=\int_0^\infty S(\omega)d\omega=\frac{1}{(2\pi)^4}\frac{4k_BT_{\circ}^2}{\rho C}\int_0^\infty\int\frac{Dk^2}{D^2k^4+\omega^2}|G({\bf k})|^2 d{\bf k}d\omega
\end{equation}
\begin{equation}\label{eq.snorm}
<|Q|^2>=\frac{1}{(2\pi)^3}\frac{k_BT_{\circ}^2}{\rho C}\int |G({\bf k})|^2 d{\bf k}=\frac{k_BT^2_{\circ}}{\rho C}\int |{\cal{E}}({\bf r})|^2 d{\bf r}.
\end{equation}
If we compare Eq. (\ref{eq.snorm}) with Eq. (\ref{eq.dt1}) we find an effective volume 
\begin{equation}\label{eq.snorm2}
\frac{1}{V_{\rm eff}}=\frac{\int|E_{01}(x,y)E_{11}(x,y)|^2dxdy}{\int |E_{01}(x,y)|^2dxdy\:\int |E_{11}(x,y)|^2dxdy}\:\frac{\pi}{L_B}.
\end{equation}
The ratio of integrals is reminiscent of the effective area for third order nonlinear processes in fibers. The easily computed $V_{\rm eff}$ is convenient in normalizing $S(\omega)$.

\begin{figure}[htbp]
\centering\includegraphics[width=12cm]{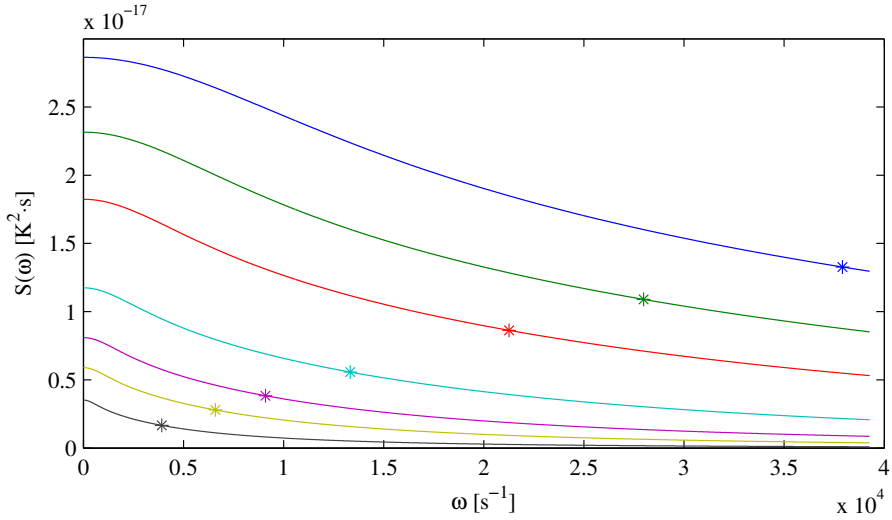}
\caption{\label{fig.sw}$S(\omega)$ at $T=300$ K for fibers with core diameters (20, 25, 30, 40, 50. 60, 80) $\mu$m (top to bottom). The star symbols indicate the frequencies of maximum STRS gain $\omega_M$.}
\end{figure}

In Fig. \ref{fig.sw} we plot $S(\omega)$ for several fibers with $NA=0.054$ and core diameters of (20, 25, 30, 40, 50, 60, 80) $\mu$m. On each curve we indicate by a star symbol the corresponding frequency of maximum gain approximated by\cite{smith3}
\begin{equation}
\omega_M=\frac{1.07\times 10^{-5}\mbox{ m$^2$/s}}{A_{\rm eff}}
\end{equation}
where $A_{\rm eff}$ is the effective area of mode LP$_{01}$. The starred values of $S(\omega)$ are used in the next section to compute spontaneous Rayleigh scattering rates.

\section{Estimate of scattered power in STRS gain band}

From Eq. (\ref{eq.fieldcouple}) we can write an equation for the time averaged power scattered into LP$_{11}$ in the frequency band $\Delta \omega$ after propagating half of a beat length,
\begin{equation}\label{eq.power}
\Delta P_{11}=\biggl[\frac{k_{\circ}\alpha L_B}{2}\biggr]^2 S(\omega)\Delta \omega \; P_{01}.
\end{equation}
For the seven fibers plotted in Fig. \ref{fig.sw}, the product
\begin{equation}
S(\omega_M)\;\omega_M\;L_B^2\approx 3\times 10^{-18} \mbox{ K$^2\cdot$m$^2$}
\end{equation}
is nearly constant at this value. If we choose a bandwidth of $\Delta \omega= 0.1\omega_M$ Eq. (\ref{eq.power}) becomes
\begin{equation}\label{eq.rateper}
\Delta P_{11}=4\times10^{-16}\:P_{01}.
\end{equation}
Over a full beat length instead of a half beat length but taking only the red shifted half of the spectrum the value remains the same. This rate is within a factor of 3 of our initial rough estimate in Section 2. We note that the rate of spontaneous thermal Rayleigh scatter into other modes is approximately proportional to their STRS gain, assuming similar beat lengths and peak gain frequencies, and those gains are generally smaller\cite{smith4} than that of LP$_{11}$. More precise comparisons require the evaluation of $S(\omega)$ in Eq. (21) using the actual modal profiles.

An exact general comparison between quantum noise and spontaneous Rayleigh scattering is not possible because quantum noise can be considered to be present at the fiber input in the stochastic electrodynamic approximation if there are no LP$_{11}$ losses in the fiber, while spontaneous Rayleigh scattering occurs along the length of the fiber and is proportional to the power in mode LP$_{01}$ and to the square of the core temperature. Nevertheless, we can make an order of magnitude comparison. The quantum noise power in the STRS gain band is
\begin{equation}
P_{QN}=\frac{h\nu\Delta\omega}{2\pi}
\end{equation}
where $\omega$ is the seed frequency and $\Delta \omega$ is the effective STRS linewidth. In both spontaneous Rayleigh scattering and quantum noise the $\Delta \omega$ of interest is the width of the STRS gain. It is independent of the seed linewidth. Using a seed wavelength of 1064 nm and a gain linewidth of 200 Hz gives $P_{QN}=4\times 10^{-17}$ W.

In modeling STRS seeding by spontaneous Rayleigh scattering, the rate of power scatter from LP$_{01}$ into LP$_{11}$ per beat length given by Eq. (\ref{eq.rateper}) can be used. This rate for a bandwidth of 0.1$\omega_M$ can be adjusted for other bandwidths by multiplying by the bandwidth ratio. The rate is also proportional to $T_{\circ}^2$, so the actual core temperature can also be taken into account. Considering a typical core temperature rise of 100 K, a contributing length of perhaps ten beat lengths, and a power of 10 W in LP$_{01}$, the spontaneously scattered seed power is perhaps 10$^2$-10$^3$ times larger than the quantum noise power. This implies the computed maximum STRS threshold based on spontaneous scatter will be reduced by approximately 10-20\% compared with that computed for quantum noise alone. To aid in estimating the degree of amplitude modulation of the seed light required to approach this threshold, we refer the reader to Fig. 3 of ref. \cite{smith5} and the associated discussion.

Finally, we note that our treatment of spontaneous thermal Rayleigh scattering is most appropriate for large diameter fibers since it ignores the thermal boundary conditions at the fiber outer diameter. This seems reasonable considering the core diameter is ususally ten or more times smaller than the full fiber diameter. It is also clear that other fibers with different guiding index profiles will have different $S(\omega)$ curves. However, since the mode instability threshold depends only logarithmically on the initial seed power, the estimates developed here should provide a useful guide for most fiber types.
\end{document}